\documentclass[conference]{IEEEtran}
\IEEEoverridecommandlockouts

\usepackage{cite}
\usepackage{amsmath,amssymb,amsfonts}
\usepackage{algorithmic}
\usepackage{graphicx}
\usepackage{textcomp}
\usepackage{xcolor}

\usepackage{multirow}
\usepackage{todonotes} 

\usepackage{xspace}
\usepackage{enumitem}
\usepackage[most]{tcolorbox}
\usepackage{varwidth}
\usepackage{comment}
\usepackage{amssymb}
\usepackage{booktabs}
\usepackage{url}

\usepackage[most]{tcolorbox}

\newtcolorbox{searchstring}[1][]{
  colback=gray!8,
  colframe=gray!55,
  fontupper=\ttfamily\scriptsize,
  boxrule=0.6pt,
  arc=2pt,
  left=1pt, right=1pt, top=2pt, bottom=2pt,
  #1
}

 \newcommand{\name}{\textsc{A3D}\xspace} 
 \newcommand{\longName}{Annotated AADL Architecture Dataset\xspace}

\def\BibTeX{{\rm B\kern-.05em{\sc i\kern-.025em b}\kern-.08em
    T\kern-.1667em\lower.7ex\hbox{E}\kern-.125emX}}
\begin{document}

\title{Mining Architectural Models in the Wild:\\ A Curated Dataset of AADL Models\\
}

\author{
\IEEEauthorblockN{Thi Dinh Tran}
\IEEEauthorblockA{\textit{Gran Sasso Science Institute}\\
L'Aquila, Italy\\
thidinh.tran@gssi.it}
\and
\IEEEauthorblockN{Maria Teresa Rossi}
\IEEEauthorblockA{\textit{Gran Sasso Science Institute}\\
L'Aquila, Italy\\
mariateresa.rossi@gssi.it}
\and
\IEEEauthorblockN{Ludovico Iovino}
\IEEEauthorblockA{\textit{Gran Sasso Science Institute}\\
L'Aquila, Italy\\
ludovico.iovino@gssi.it}
\and
\IEEEauthorblockN{Leonardo Mariani}
\IEEEauthorblockA{\textit{University of Milano-Bicocca}\\
Milan, Italy\\
leonardo.mariani@unimib.it}
}

\maketitle

\begin{abstract}

Empirical research on software architecture maintenance and evolution is hindered by the scarcity of large-scale, validated, and richly annotated model datasets. Existing benchmarks are limited in scale, provide only model-level annotations, or lack alignment with reference architectures, thereby restricting reproducible studies of architectural reuse, conformance, and AI-driven analysis. To address this gap, we present \name, a curated dataset of 1,115 Architecture Analysis and Design Language (AADL) models mined from public GitHub repositories and validated with the OSATE tool. The dataset spans 7 generic domains and 52 specific-problem clusters, and provides two annotation layers: (i) domain- and problem-level labels, which capture different model aggregation criteria, and (ii) component-level mappings to three reference architectures, which have been externally validated through a user study. Models are released in AADL and JSON formats with full provenance, a reproducible mining pipeline, and the validation protocol.  

\end{abstract}


\section{Introduction}
Software architecture models play a central role in the design, analysis, and evolution of software systems. These models are frequently reused, analyzed, and evolved throughout the system life-cycle, making them valuable artifacts for empirical studies on architecture evolution, architectural reuse, and automated architecture analysis~\cite{hasselbring2018software}. Among the modeling languages adopted in these contexts, the Architecture Analysis and Design Language (AADL)~\cite{AADL} has gained significant adoption due to its precise semantics and support for rigorous analyses of system structure, behavior, and resource usage~\cite{besnard2017formal,tan2022comprehensive}. According to~\cite{malavolta2012industry}, AADL is among the most widely used architectural languages in industry\cite{zou2025automated}, alongside ArchiMate~\cite{josey2016introduction} and industry-specific notations. 

Despite this growing interest, publicly available datasets of architectural models remain limited in both scope and annotation richness. 
The ArDoCo dataset~\cite{fuchss2022establishing} focuses on traceability link recovery between software architecture documentation and models, providing a small number of projects with manually curated trace links. 
The EA ModelSet dataset~\cite{glaser2025extended} offers a larger collection of ArchiMate enterprise architecture models gathered from multiple sources; however, its annotations remain high-level. Similarly, ModelSet~\cite{lopez2024modelset} introduces a large dataset of software models designed for machine learning tasks. 
Overall, as summarized in Table~\ref{tab:dataset_comparison}, existing datasets address specific tasks and typically define annotations at the model level, not capturing the roles and relationships of individual architectural components. As a result, tasks such as architectural clustering, similarity analysis, and the identification of reusable architectural solutions remain challenging, as they require both meaningful contextual annotations and fine-grained structural information across models~\cite{krishnakumar2023domain}.


\begin{table*}[t]
\centering
\caption{Comparison of publicly available software architecture model datasets across key dimensions for architectural analysis.}
\label{tab:dataset_comparison}

\resizebox{\textwidth}{!}{
\begin{tabular}
{
p{3cm}
p{1.5cm}
p{1.2cm}
p{4cm}
p{4cm}
p{4.2cm}
p{1.4cm}
p{4cm}
}

\toprule
\textbf{Dataset} 
& \textbf{Modeling Language} 
& \textbf{\#Models} 
& \textbf{Quality Assurance Method} 
& \textbf{Annotation Granularity} 
& \textbf{Structural Representation} 
& \textbf{RA Alignment} 
& \textbf{Target Research Tasks} \\
\midrule

ArDoCo Benchmark ~\cite{fuchss2022establishing}
& PCM/UML 
& 5 projects 
& Manual curation
& Traceability links (model–doc) 
& Implicit (model structure only) 
& No 
& Traceability link recovery \\

EA ModelSet ~\cite{glaser2023ea}
& ArchiMate 
& 863 
& Format standardization 
& Model-level labels (limited domain info) 
& Implicit (graph-based models) 
& No 
& Enterprise architecture analysis, model retrieval \\

Extended EA ModelSet ~\cite{glaser2025extended}
& ArchiMate 
& 977 
& Manual curation 
& Model-level labels (limited domain info) 
& Implicit (graph-based models) 
& No 
& Architecture analysis, ML applications \\

ModelSet ~\cite{lopez2024modelset}
& UML/Ecore 
& 10k+ 
& Manual curation 
& Model-level labels (for ML tasks)
& Implicit (metamodel-based structure) 
& No 
& Model classification, clustering \\

\midrule
\textbf{A3D} 
& \textbf{AADL} 
& \textbf{1,115} 
& \textbf{Automated syntactic validation (OSATE) + manual review} 
& \textbf{Multi-level: domain, problem, and component-level mappings} 
& \textbf{Explicit (component–connection structure)} 
& \textbf{Yes} 
& \textbf{Architectural clustering, reference architecture conformance, ML/LLM applications} \\

\bottomrule
\end{tabular}
}
\vspace{0.5em}
\end{table*}

To address this gap, we present the \longName (\name), a curated dataset of 1,115 validated AADL models mined from GitHub. The models are automatically collected and systematically validated using the OSATE tool~\cite{OSATE} to ensure syntactic correctness and conformance with the AADL metamodel. \name advances the state of the art along three key dimensions: i) \emph{scale}, since it is the largest curated dataset of AADL models, and among the largest in general, ii) \emph{multi-level domain-sensitive annotation}, since it includes two complementary annotation layers that represent how models can be used to address problems defined at different abstraction levels, and iii) \emph{architectural context annotation}, which defines consistency relations between architectural models and reference architectures (RAs). To the best of our knowledge, this is the first dataset that combines these characteristics at this scale.
Given the diversity of architectural models across domains and problem types, \name establishes a reusable dataset for empirical studies on software architecture analysis and evolution, including AI-driven approaches.

In a nutshell, the contributions of this work are threefold:
\begin{itemize}[leftmargin=*]
\item \name, a large-scale dataset of 1,115 validated AADL architectural models mined from GitHub repositories, accompanied by full provenance metadata and a reproducible data collection pipeline.
\item A multi-level annotation scheme including generic domain- and specific problem-level labels that enable architectural clustering, similarity analysis, and comparative studies across architectural models.
\item A novel component-level alignment between architectural models and RAs, externally validated through a user study, enabling empirical research on architectural reuse, RA conformance, and automated architecture alignment.
\end{itemize}

\section{Dataset Construction}

\begin{figure*}[t]
\centering
\includegraphics[width=0.7\textwidth]{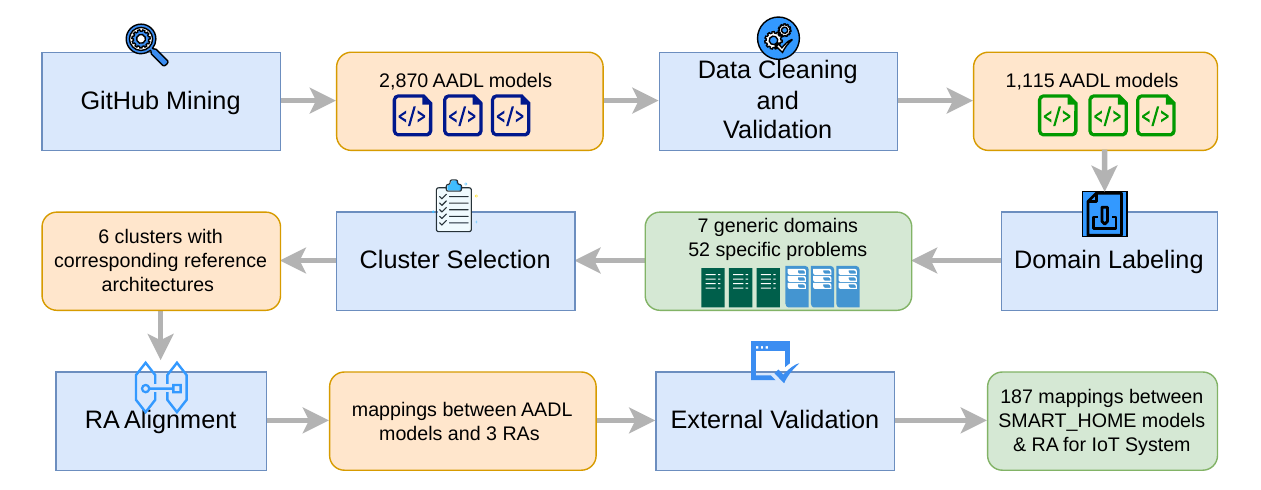}
\caption{Overview of the dataset construction and annotation pipeline.}
\label{fig:pipeline}
\end{figure*}

Figure~\ref{fig:pipeline} provides an overview of the dataset construction and annotation pipeline. 
To establish a large-scale corpus of architectural models, we mined public GitHub repositories containing AADL specifications. We performed a GitHub code search, via the GitHub REST API (v3), using the query:
\begin{searchstring}
`(extension:aadl \textbf{OR} extension:aadl2 \textbf{OR} extension:aaxl2)` 
\end{searchstring}
The search was conducted in March 2025 and exhaustively paginated to retrieve all matching repositories. Forked repositories were excluded to reduce redundancy. 

For each selected repository, we cloned the default branch at its latest commit at the time of collection. Architectural model files were extracted as individual units of analysis; i.e., each AADL file is treated as a single model. 
Provenance metadata, including repository URL, file path, and license information, were preserved. Duplicate files were identified and removed using SHA-256 content hashing. This process resulted in an initial collection of 2,870 distinct AADL models.

\begin{figure}[!htp]
    \centering
    \includegraphics[width=\columnwidth]{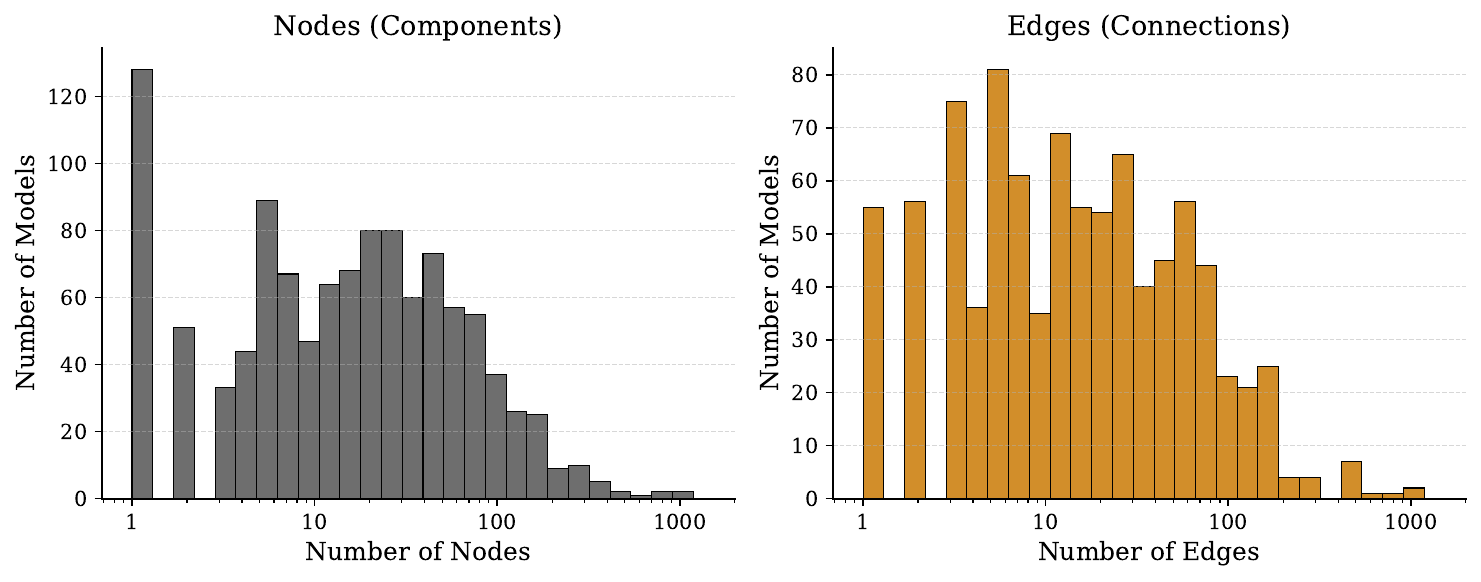}
    \caption{The distribution of model sizes.}
    \label{fig:model_size_distribution}
\end{figure}

To ensure dataset quality and suitability for empirical research, all extracted models were validated using the OSATE validator, which performs automated checks to verify that the AADL models are structurally sound, semantically correct, and compliant with the standard. Models exhibiting syntactic errors, unresolved dependencies, or inconsistencies with respect to the AADL metamodel were excluded. After validation and filtering, 1,115 models remained in the final dataset. This guarantees that all models are syntactically correct and compatible with standard AADL tooling.
The resulting dataset contains models of varying size and structural complexity. On average, models include 29.4 components (nodes) and 34.0 connections (edges). As shown in Figure~\ref{fig:model_size_distribution}, the distributions of both nodes and edges are highly right-skewed, with the majority of models containing fewer than 50 components and connections. At the same time, the dataset exhibits a long-tailed behavior, with a small number of significantly larger models. In particular, the largest model contains 934 nodes and 1,188 edges, highlighting the presence of extreme cases. The similarity of the two distributions suggests that the number of connections scales proportionally with the number of components.

The models originate from repositories spanning multiple application sectors, including aerospace, healthcare, automotive, and other embedded and cyber-physical domains, based on repository metadata and documentation.
\section{Annotation Methodology}

To support different architectural analysis scenarios, we constructed complementary annotations of the collected models: models labeled according to the domain they belong to, according to the specific problem they address, and, for a selection of models, according to the reference architecture they fulfill.   

\subsection{Generic Domain- and Specific Problem-Level Labeling}

We classified models according to two criteria reflecting common architectural analysis scenarios: clustering models by \emph{generic domain}, to support the exploration of architectures within a broad sector (e.g., avionics, automotive, etc.), and clustering models by \emph{specific problem}, to identify architectures addressing the same concrete design problem and enable targeted reuse (e.g., drone control, flight management system, etc.). 
The domain taxonomy follows a bottom-up approach derived from the dataset itself rather than adopting a predefined taxonomy. 

Two authors independently inspected and classified the models. In each annotation round, both reviewers independently examined the models and assigned labels based on their interpretations of the architecture and the accompanying repository documentation. After the independent phase, the reviewers met to compare their classifications and discuss any discrepancies. Ambiguous cases were resolved collaboratively through discussion and refinement of the labeling criteria. Over multiple rounds, this iterative process led to a stable consensus on the final ground truth. To quantify the level of agreement between annotators before reconciliation, we computed \emph{Cohen's Kappa} coefficient~\cite{artstein2017inter}, resulting in $\kappa=0.764$, indicating substantial agreement.

The set of \emph{generic domains} emerged incrementally during the annotation process and was finalized during the consensus phase. In total, seven recurrent domains were identified, covering a diverse range of contexts, including aerospace, automotive, smart building, and related domains. 

\begin{figure}[!htp]
    \centering
    \includegraphics[width=1\columnwidth]{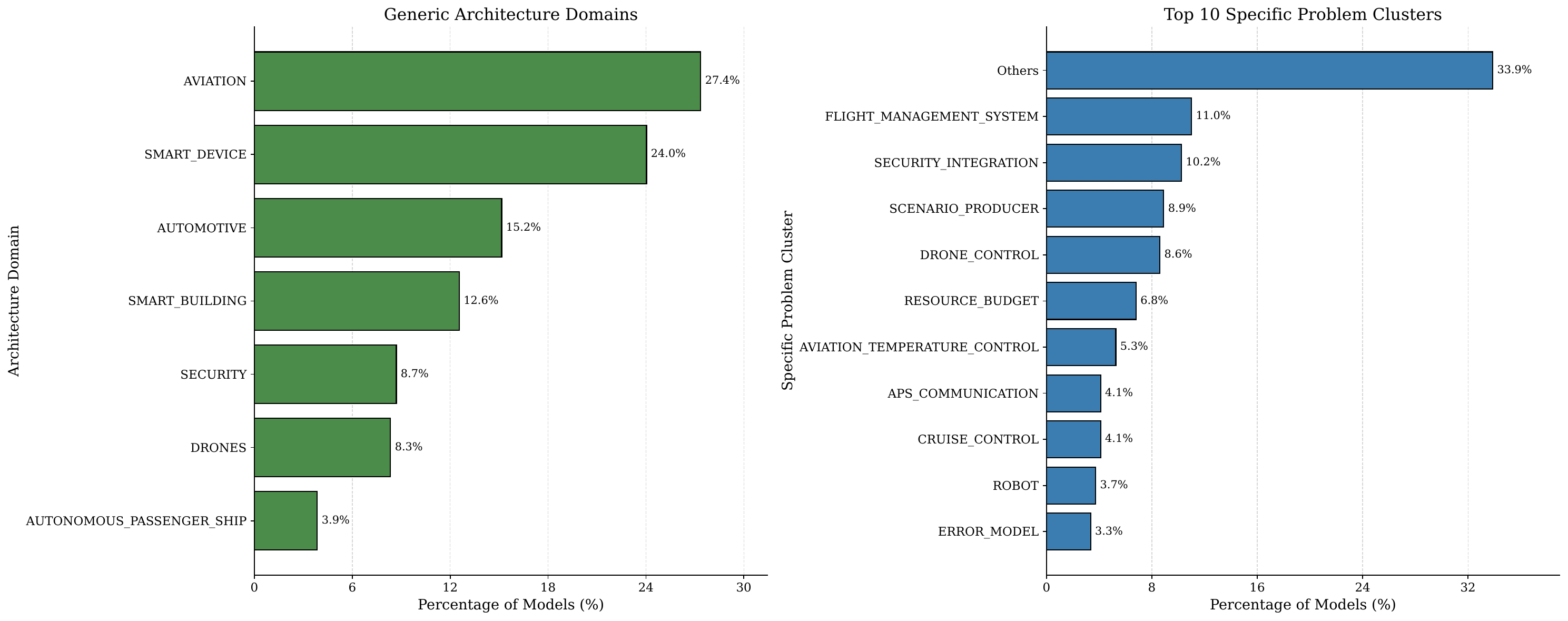}
    \caption{Distribution of models across generic domains (left) and the top 10 specific problem clusters (right).}
    \label{fig:application_specific_domain_distribution}
\end{figure}
Figure~\ref{fig:application_specific_domain_distribution} reports the distribution of models across both the generic architecture domains (left) and the top 10 specific problem clusters (right). The left panel shows that \texttt{AVIATION} and \texttt{SMART\_DEVICE} account for the largest proportions (27.4\% and 24.0\%, respectively), followed by \texttt{AUTOMOTIVE} (15.2\%) and \texttt{SMART\_BUILDING} (12.6\%). The remaining domains, including \texttt{SECURITY}, \texttt{DRONES}, and \texttt{AUTONOMOUS\_PASSENGER\_SHIP}, contribute smaller shares. The distribution is moderately imbalanced but not dominated by a single domain, indicating a diverse dataset that covers a range of application contexts.

To derive the ground truth for the \emph{specific problem} criterion, we analyzed the system-level objective and functionality represented in each architecture. In this context, a \emph{specific problem} denotes the concrete design objective addressed by the architecture, independently of the generic domain in which it is deployed. Using the same review and consensus process adopted for domain labeling, we obtained 52 clusters representing distinct architectural problems, with $\kappa=0.893$, denoting strong agreement among annotators.

The right panel of Figure~\ref{fig:application_specific_domain_distribution} 
presents the distribution of the top 10 most frequent problem clusters, with the remaining clusters aggregated into the \texttt{OTHERS} category. The figure reveals a clear long-tailed distribution: while a few recurrent problems such as \texttt{FLIGHT\_MANAGEMENT\_SYSTEM} (11.0\%) and \texttt{SECURITY\_INTEGRATION} (10.2\%) account for a notable portion of the dataset, a substantial fraction (33.9\%) is composed of less frequent or highly specialized problems grouped under \texttt{OTHERS}. This indicates that, unlike the generic domain-level categorization, the specific problem-level space is significantly more fine-grained and diverse. Moreover, the relatively balanced proportions among the top clusters (ranging from 3\% to 11\%) suggest that no single problem dominates the dataset.

\begin{figure}[!htp]
    \centering
    \includegraphics[width=\columnwidth]{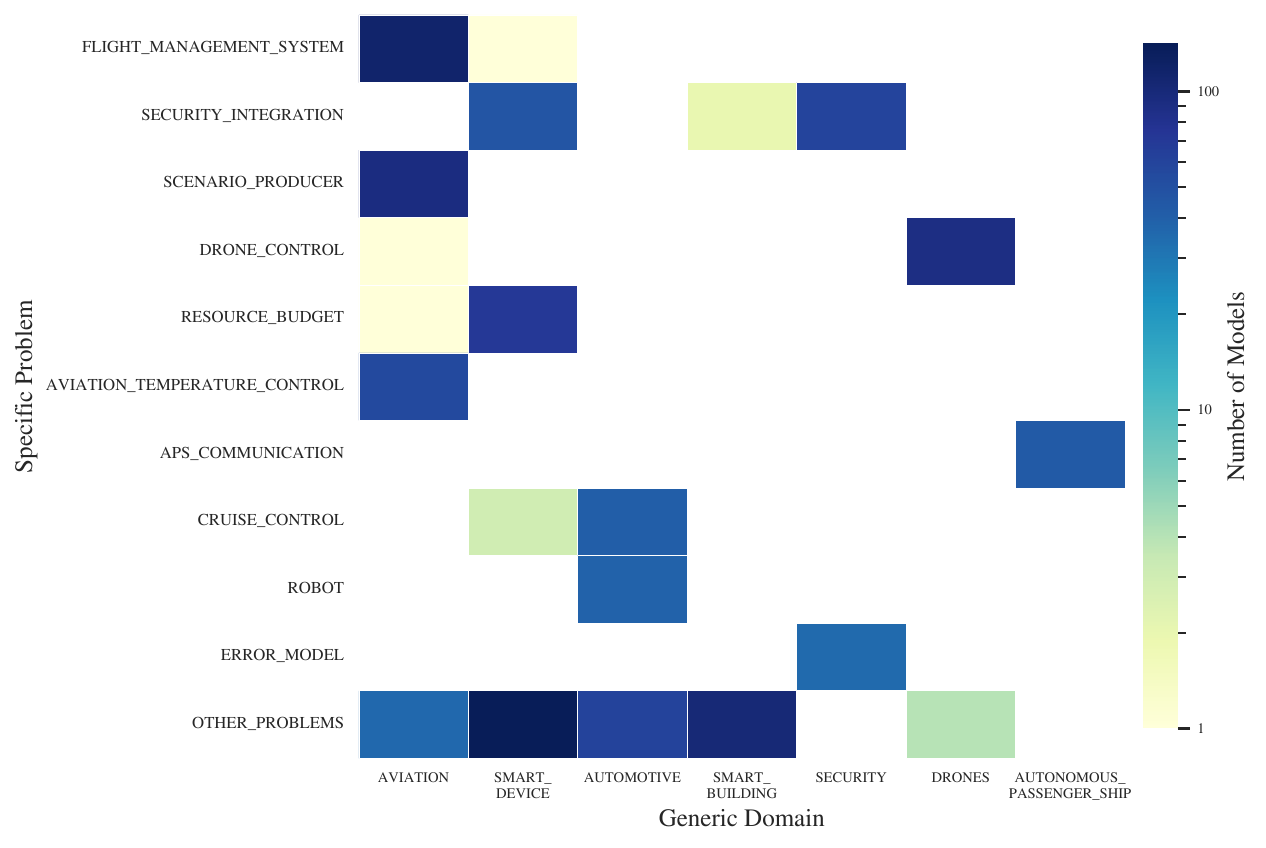}
    \caption{Relationship Between Generic Domains and Specific Problem.}
    \label{fig:heatmap}
\end{figure}

Figure~\ref{fig:heatmap} illustrates the relationship between generic domains and the most frequent specific problem clusters. The distribution reveals that several clusters are strongly associated with a single domain (e.g., \texttt{FLIGHT\_MANAGEMENT\_SYSTEM} in \texttt{AVIATION} and \texttt{DRONE\_CONTROL} in \texttt{DRONES}), indicating domain-specific architectural patterns. At the same time, other clusters (e.g., \texttt{SECURITY\_INTEGRATION} and \texttt{SMART\_BUILDING}) span multiple domains, reflecting the presence of cross-domain architectural solutions. This heterogeneous distribution highlights that generic domain-level categorization alone is insufficient to capture architectural similarity, motivating the need for complementary problem-level annotations.


\subsection{Component-Level Reference Architecture Alignment}

While the generic domain and specific problem annotations enable clustering-based analyses, they do not capture the structural correspondence between architectural models and RAs commonly used in their respective domains. To support research on architectural reuse and conformance, we introduce a second annotation layer that aligns architectural components with the key elements of corresponding RAs.

We first identified 6 clusters derived from the \emph{specific problem} grouping that contain at least ten models and exhibit a high degree of architectural commonality. Such clusters are more likely to reflect recurring architectural solutions and therefore provide an appropriate basis for RA identification. Among the candidate clusters, we selected the \texttt{SMART\_HOME} cluster, which contains 25 AADL models and represents IoT-based smart home systems. Due to the high manual effort required for component-level alignment and external validation, we focused on this cluster as a primary case study to establish and rigorously validate the annotation methodology.

\begin{figure}[!htp]
    \centering
    \includegraphics[width=\columnwidth]{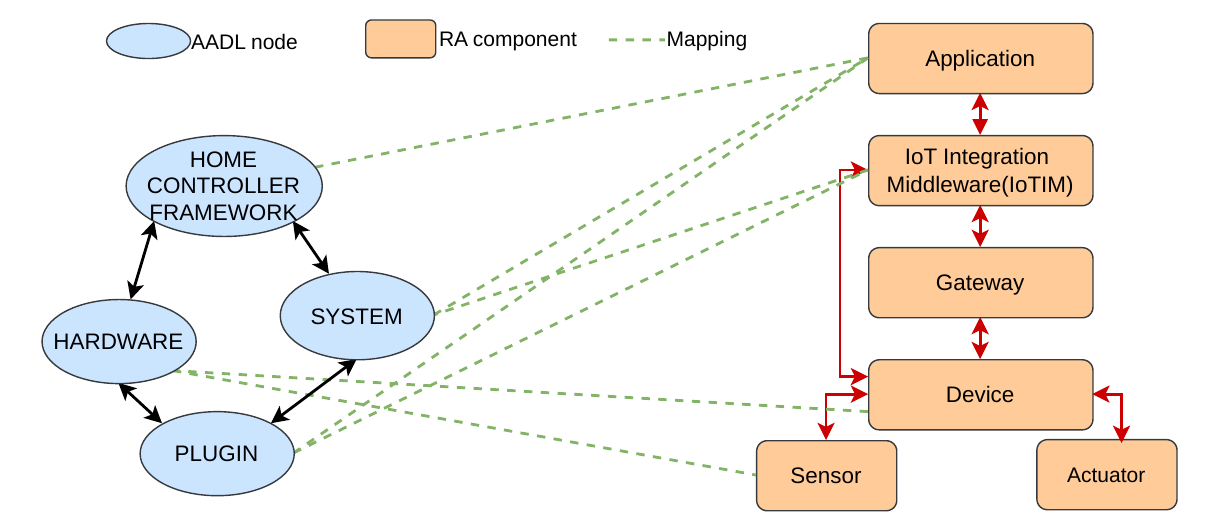}
    \caption{Example of aligning AADL components (left) with IoT RA elements (right).}
    \label{RA_alignment_example}
\end{figure}

For this cluster, the corresponding RA was identified as the \emph{IoT Reference Architecture}\cite{di2023mastering}. This choice is motivated by the architectural structure of the \texttt{SMART\_HOME} models, which typically implement IoT-based systems composed of sensing devices, communication middleware, data processing services, and application-level control components. 
The author team provided an initial alignment by manually mapping the components in each AADL model to the corresponding elements of the IoT RA. Mappings are established based on the functional responsibility of AADL components and their role within the architectural data flow, rather than naming similarity alone. Annotators consider component behavior, interactions, and graphical illustrations, and may assign one-to-many mappings when a component spans multiple RA roles. For instance, Figure~\ref{RA_alignment_example} shows a representative alignment for a smart home sensor unit model to its IoT RA. The complete annotation guidelines, including definitions and examples, are publicly available in the dataset repository to facilitate reproducibility.

We validated the mappings externally with 12 researchers experienced in software modeling and RAs. The 187 mappings were split into four batches. Each participant evaluated one batch, yielding three independent judgments per mapping. For each item, participants were provided with the AADL model, the component, the proposed RA element, and a link to the graphical mapping. Additional supporting material included descriptions of all RA components and a visualization of the IoT RA. Final labels were assigned by majority voting; high-disagreement cases were flagged and resolved by the authors.

The external validation results indicate substantial agreement among participants, with only 10.2\% 
of the mappings exhibiting disagreement among participants. 

In addition to the \texttt{SMART\_HOME} cluster, we also applied the same component-level alignment process to two other clusters, namely \texttt{SELF\_DRIVING\_CAR}\cite{velasco2020autonomous} and \texttt{SMART\_PARKING}\cite{ozkaya2025reference}. These annotations demonstrate the applicability and generalizability of the proposed alignment methodology across multiple domains with similar architectural patterns. 

Beyond the manually established mappings, we further corroborated the structural alignment by formally verifying the conformance of the annotated AADL models against their respective RAs using an automatic conformance-checking tool~\cite{bucaioni2023reference}. 
Applied across all three annotated clusters, \texttt{SMART\_HOME}, \texttt{SELF\_DRIVING\_CAR}, and \texttt{SMART\_PARKING}, the tool confirmed that all the mapped AADL models in each domain conform to their corresponding RAs. This formal verification strengthens the evidence about the correctness of the established mappings.        

\section{Dataset Structure and Availability}



\emph{Dataset Organization.} The dataset is organized in three folders. 

The \texttt{models/} folder contains the 1,115 validated AADL architectural models. Each model is treated as an independent unit of analysis and is provided in two formats: \texttt{.aaxl2} (\texttt{xmi/}) and normalized JSON (\texttt{json/}) to support automated processing and machine learning applications.

The \texttt{reference\_architectures/} folder contains the RAs used for component-level alignment. For each RA, the \texttt{definition/} folder provides its structural specification (e.g., \texttt{.ecore}, \texttt{.gramm}) together with a description of its key components. The \texttt{mappings/} folder includes both machine-readable mappings between AADL components and RA elements (\texttt{json/}) and visual representations (\texttt{visual/}) used during validation. The \texttt{references/} folder contains supporting literature describing the RA.

The \texttt{annotations} folder contains files with annotations that define how models are clustered and the mapping between models and RAs. Visual representations of the mappings are also included in the folders.

\emph{Metadata Schema.} A single metadata file lists, for each model, its name, source repository, file path, component/connection counts, and annotated labels. The \texttt{schema/} folder provides the JSON schemas of both the normalized model representation and the component mapping files.

\emph{Dataset Availability.} To support open science and reproducible research, the dataset (v1.0.0, CC-BY-4.0) is publicly released~\cite{replication-package} with an immutable Zenodo snapshot (DOI: 10.5281/zenodo.19819886).

To facilitate reproducibility, the repository also provides scripts implementing the full data collection pipeline described in Section~II, including mining, preprocessing, and validation steps. These scripts allow users to regenerate the dataset or extend it with new models. 
The experimental protocol used for collecting human validation judgments is also included to ensure transparency and replicability of the annotation process.
\section{Research Opportunities}

The proposed dataset enables a range of research directions in software architecture analysis, maintenance, and evolution.

\noindent \textbf{Architectural clustering and similarity analysis.} 
The availability of domain- and problem-level annotations enables the evaluation of approaches for clustering and detecting similarities among architectural models, fostering reuse, as recently demonstrated in~\cite{tran2026structuralsemantic}.

\noindent \textbf{AI-driven techniques for architectural models.} 
The dataset provides normalized JSON representations and labeled data that enable machine learning and LLM applications, including classification and semantic similarity estimation across architectural models.

\noindent \textbf{RA reuse and conformance analysis.} 
The mapping between AADL models and RAs supports research on architectural reuse and conformance: researchers can analyze how concrete implementations align with conceptual architectures and develop automated approaches for model--RA alignment.

\noindent \textbf{Architectural variability and design diversity.} 
By collecting models from heterogeneous repositories and domains, the dataset enables studies on recurring architectural patterns, variability across implementations of the same problem, and the impact of architectural design decisions on system structure.

\section{Limitations}

\noindent Despite its scale and richness, the dataset presents several limitations that should be considered when using it.

\noindent \textbf{The dataset may reflect biases from publicly available repositories.} 
Since the models are mined from GitHub, the dataset may be biased toward open-source projects with varying levels of maturity, documentation quality, and domain coverage, potentially underrepresenting certain application areas.

\noindent \textbf{Architectural models are analyzed at the file level.} 
Each AADL file is treated as an independent model, which may overlook cross-file dependencies in multi-file projects. This aspect may reflect models that were removed due to missing validation. 

\noindent \textbf{Annotation remains partially subjective despite validation.} 
Despite independent review and consensus, the interpretation of architectural intent may remain ambiguous in borderline cases.

\noindent \textbf{RA alignment is limited to a subset of the dataset.} 
The RA alignment has been applied to a subset of clusters, namely \texttt{SMART\_HOME}, which was also externally validated, \texttt{SELF\_DRIVING\_CAR}, and \texttt{SMART\_PARKING}. 



Addressing these limitations represents an important direction for the future extension of this dataset.
\section{Conclusion}

We presented A3D, a curated dataset of 1,115 validated AADL models combining multi-level annotations, plus a reproducible mining pipeline and rich provenance metadata. Released under CC-BY-4.0, A3D is a reusable resource for empirical research on architectural analysis, design, and evolution, and for AI-driven approaches leveraging architectural knowledge.

\bibliographystyle{IEEEtran}
\bibliography{references}

\end{document}